\documentclass[conference]{IEEEtran}

\usepackage{amsmath, amssymb} 
\usepackage{enumitem} 
\usepackage{calc} 
\usepackage{array} 
\usepackage{cite} 
\usepackage{hyperref} 
\usepackage{bm} 
\usepackage{booktabs} 
\usepackage{graphicx} 
\usepackage{float} 
\usepackage{tabularx} 
\usepackage{multirow}
\usepackage{url}

\begin{document}

\title{PROMETHEE-based Modeling of Endogenous Behavioral Uncertainty of EV Owners }

\author{\IEEEauthorblockN{Dipayan Sarkar}
\IEEEauthorblockA{Electrical and Computer Engineering\\
University of Central Florida\\
Orlando, USA\\
dipayan.sarkar@ucf.edu}
\and
\IEEEauthorblockN{Qifeng Li}
\IEEEauthorblockA{Electrical and Computer Engineering\\
University of Central Florida\\
Orlando, USA\\
Qifeng.Li@ucf.edu}
}

\maketitle

\begin{abstract}
The electric vehicle (EV) charging demands (CD) are jointly determined by the EV owners' behavior (i.e., human factor) and the electricity prices (i.e., decisions of distribution system operators (DSO)). However, most existing studies either neglect the decision-dependent nature of EVCD uncertainty or idealistically treat EV owners as perfect decision-makers. This paper formulates the optimal operation of power distribution systems (PDS) as a distributionally robust chance-constrained (DRCC) problem considering EVCDs as endogenous uncertainty (i.e., decision-dependent uncertainty). The Preference Ranking Organization Method for Enrichment Evaluation (PROMETHEE) is introduced to capture the human factor of EV owners in the proposed ambiguity set. Case studies on IEEE test systems demonstrate that the proposed method achieves superior performance compared to deterministic and conventional DRCC approaches, thereby enhancing resilience and security in PDS operations.
\end{abstract}


\IEEEpeerreviewmaketitle

\section{Introduction}
To ensure optimal operation of distribution systems, distribution system operators (DSOs) must account for various sources of uncertainty \cite{li2021uncertainty}. These uncertainties are broadly categorized as exogenous uncertainty (ExU), originating from uncontrollable factors such as variability in renewable energy sources (RES), and endogenous uncertainty (EnU), that is affected by operator decisions, such as electric vehicle (EV) charging demand (CD). While the stochastic nature of RES has been extensively investigated in existing literature \cite{singh2022uncertainty}, the uncertainty associated with EVCD remains underexplored. Given that global EV charging demand is projected to reach nearly 700 TWh by 2030 \cite{IEA}, overlooking these behavioral-driven uncertainties risks grid stress, inefficient resource utilization, and compromised reliability. 

Recent research has increasingly addressed endogenous uncertainty (EnU) within power systems, examining its impact on demand response \cite{bayat2023stochastic}, reserve deployment \cite{zhang2021robust}, and EVCD \cite{wang2024v2g, wang2025robust, yi2024optimal}. Despite these advancements, existing EVCD models remain limited in scope. Previous studies adopted deterministic frameworks with minimal coordination between transportation system operators and DSOs \cite{sharma2024enhanced}. Probabilistic approaches such as modelling EVCD using Gaussian mixture models \cite{shi2022day} or lognormal distributions \cite{shi2020integration} were also used considering EVCD as ExU. Other studies considered EVCD as EnU arising from pricing strategies \cite{wang2024v2g, wang2025robust} or congestion effects \cite{yi2024optimal}, but commonly assumed rational behavior among EV owners, optimizing decisions solely based on charging costs or waiting times.   

The assumption of rational decision-making among EV owners presents a significant limitation in capturing the human factor. Existing studies predominantly assume that EV users behave in an economically optimal manner \cite{wang2024v2g, wang2025robust}. However, in practice, human behavior is often influenced by cognitive biases \cite{tversky1974judgment}, including habitual selection of charging stations and resistance to change that leads to deviation from an optimal solution. These behavioral patterns critically affect the spatial distribution of EVCD, particularly as the density of EVs and charging infrastructure increases. Failure to incorporate such behavioral heterogeneity may result in underestimation of demand variability, leading to inefficient resource allocation, localized congestion, and potential grid instability. Prevailing models either oversimplify EVCD as an exogenous input or treat it as endogenous under a singular rationality framework, thereby overlooking the human-centric dimension of uncertainty.  

To address these challenges, this paper introduces the \textit{Preference Ranking Organization Method for Enrichment Evaluation (PROMETHEE)} \cite{brans1986select}, an outranking-based multi-criteria decision-making method, to explicitly model the heterogeneous decision-making behavior of EV owners. PROMETHEE offers a structured framework for evaluating alternatives across multiple, potentially conflicting criteria. In our context, rational criteria include charging cost, and congestion duration, whereas psychological criteria encompass behavioral tendencies such as habitual bias and aversion to switching charging stations. By incorporating PROMETHEE into the EVCD modeling framework, this study captures the endogenous nature of demand with greater fidelity. The proposed approach facilitates a nuanced understanding of the interaction between EV owners’ diverse preferences and DSO pricing strategies, thereby enabling a more accurate prediction of EVCD dispersion across the distribution network. 

The rest of the paper is oganized as follows. Section II introduces the PROMETHEE method and models EV owner behavior as endogenous uncertainty to the DSO. In section III, we show distributionally robust chance-constrained (DRCC) based distribution optimal power flow (DOPF) formulation and conic reformulations of the chance constraints. Section IV covers the case study and post-optimization analysis. Finally, section V provides conclusion and outline for future work.  

\section{Proposed PROMETHEE-based Modeling of EV Owner Behavior} 
The PROMETHEE method (PM) is a widely adopted decision-making framework utilized across diverse domains, including business, healthcare, education, and resource management, where decisions must reconcile technical, economic, and behavioral dimensions \cite{behzadian2010promethee}. Its principal advantage lies in its ability to identify alternatives that best align with the decision maker’s objectives and subjective interpretation of the problem. Rather than seeking a globally optimal solution, it reflects individual preferences and priorities—an approach well-suited to modeling EV owner behavior, where decisions are often influenced by personal habits and perceptions rather than purely system-level efficiency. PROMETHEE operationalizes decision-making by quantifying pairwise preferences through preference functions, which evaluate the degree to which one alternative is preferred over another for each criterion. In this section, we first leverage PROMETHEE to model the EV owner behavior and, then, formulate this behavior model as endogenous uncertainty to the decision-making of DSO.
\subsection{Modeling EV owner behavior with PROMETHEE}
 Utilizing the V-shaped preference function with defined indifference and strict preference thresholds \cite{lolli2022decision}, the following formulation is proposed to model the heterogeneous decision behavior of EV owners (Throughout the paper, \textbf{bold} symbols denote vectors/matrices of the corresponding variables):   
\begin{IEEEeqnarray}{l}
    \mathcal{U} = \{\lambda, t_m, b\} \IEEEyesnumber\IEEEyessubnumber* \label{eq:pm1a} \\
    \Delta_j(i,k) = g_j(i) - g_j(k), \quad \forall j \in \mathcal{U}, \forall i,k \in N_{evcs} \label{eq:pm1b} \\
    \psi_{j,y}(i,k) = 
    \begin{cases}
    0 & \text{if } \Delta_j(i,k) \leq \tau^q_{j,y} \\
    \frac{\Delta_j(i,k)-\tau^q_{j,y}}{\tau^p_{j,y}-\tau^q_{j,y}} & \text{if } \tau^q_{j,y}\leq \Delta_j(i,k) \leq \tau^p_{j,y} \\ 
    1 & \text{if } \Delta_j(i,k) \geq \tau^p_{j,y}
    \end{cases}, \notag \\ \qquad \qquad \forall j \in \mathcal{U}, \forall y \in \mathcal{Z}_i, \forall i \in N_{evcs} \IEEEeqnarraynumspace \label{eq:pm1c} \\
    \pi_y(i,k) = \sum_j \psi_j(i,k) w_{y,j}^{pm}, \quad 
    \sum_j w_{y,j}^{pm} = 1, \forall y \in \mathcal{Z}_i \label{eq:pm1d} \\
    \phi^+_{i,y} = \frac{1}{|N_{evcs}|-1} \sum_{k \in N_{evcs}} \pi(i,k),  \forall y \in \mathcal{Z}_i, \forall i \in N_{evcs} \IEEEeqnarraynumspace \label{eq:pm1e} \\
    \phi^-_{i,y} = \frac{1}{|N_{evcs}|-1} \sum_{k \in N_{evcs}} \pi(k,i), \forall y \in \mathcal{Z}_i, \forall i \in N_{evcs} \IEEEeqnarraynumspace \label{eq:pm1f} \\ 
    \phi_{i,y} = \phi^+_{i,y} - \phi^-_{i,y} \quad \forall y \in \mathcal{Z}_i, \forall i \in N_{evcs} \label{eq:pm1g} \\ 
    \delta_{y,i} = 
    \begin{cases}
    1, & \text{if } i = \arg \max_{i \in N_{evcs}} \phi_{\mathrm{i,y}}, \\
    0, & \text{otherwise.}
    \end{cases}, \forall y \in \mathcal{Z}_i \label{eq:pm1h} \\ 
    n_i = \sum_{y \in \mathcal{Z}_i} \delta_{y,i}, \qquad \forall i \in N_{evcs} \label{eq:pm1i} \\
    p_{i}^{ev} = n_i Ch_{avg} \quad \forall i \in N_{evcs}, \label{eq:pm1j} \\
    \mathbf{p}^{ev} = 
    \begin{bmatrix}
        p_{1}^{ev},p_{2}^{ev},\cdots, p_{|N_{evcs}|}^{ev}
    \end{bmatrix}, \IEEEyessubnumber* \label{eq:pm1k}
\end{IEEEeqnarray}
where $\mathcal{U}$ is the set of criteria: charging price $\lambda$, congestion time $t_{m}$ and owner specific bias $b$. Here, $N_{evcs}, Z_i$ indicate the set of EV charging stations (EVCSs) and set of EV owners in \textit{i}th EVCS. While $g_j(i), g_j(k)$ indicates the performance of \textit{i}th and \textit{k}th EVCS for \textit{j}th criterion, $\Delta_j(i,k)$ indicates the difference between them. The strict preference and indifference is denoted by $\tau^p, \tau^q$ and $\psi, \pi$ denotes preference function and preference index. The weights of the criteria is marked by $w$. $\phi^+, \phi^-, \phi$ refers to positive, negative, and net outranking flow respectively. A binary variable $\delta$ is used for rerouting decisions, and $CH_{avg}$ indicates the average charge for each EV.  

Equation \eqref{eq:pm1b} quantifies the criterion-wise difference between alternatives, while the V-shaped preference function defined in \eqref{eq:pm1c} determines the degree of preference between two options. Subsequently, \eqref{eq:pm1e}, \eqref{eq:pm1f}, and \eqref{eq:pm1g} compute the positive, negative, and net outranking flows, respectively. Equation \eqref{eq:pm1h} indicates the switching decision of EV owners, \eqref{eq:pm1i} and \eqref{eq:pm1j} calculates the redistributed EVs and EVCD at \textit{i}th EVCS. This formulation effectively captures the heterogeneity in EV owners’ prioritization of charging cost, waiting time, and habitual biases during EVCS selection. 

The PM model treats each EVCS as a competing alternative, with rerouting decisions evaluated across these criteria: bias factor $b$, charging cost $\lambda$, and congestion time $t_m$. To reflect the diversity in EV owner preferences, random weights $w$ are assigned to these criteria and normalized to ensure their sum equals one. Additionally, individual variability is captured by randomly assigning strict preference  $\tau^p$ and indifference $\tau^q$ thresholds for each EV owner. The PM model produces net outranking flows for each alternative charging stations. If the net outranking flow of the current EVCS is not the highest, an EV owner will switch to the EVCS with the highest score. 

\subsection{EV Owner Behavior as Endogenous Uncertainty to DSO}
 To this end, we calculate the mean and covariance using the $\mathbf{p}^{ev}$ from \eqref{eq:pm1k} for $K^{sc}$ number of scenarios. 
\begin{IEEEeqnarray}{rCl}
    \bar{\mathbf{p}}^{ev} &=& 
    \frac{1}{|K^{sc}|} \sum_{k=1}^{K^{sc}} \mathbf{p}_{k}^{ev} \IEEEyesnumber\IEEEyessubnumber* \label{eq:amb2a} \\
    \boldsymbol{\Sigma} &=& 
    \frac{1}{|K^{sc}|}\sum_{k=1}^{K^{sc}}
    \bigl(\mathbf{p}_{k}^{ev}-\bar{\mathbf{p}}^{ev}\bigr)
    \bigl(\mathbf{p}_{k}^{ev}-\bar{\mathbf{p}}^{ev}\bigr)^{T} \label{eq:amb2b}
\end{IEEEeqnarray}
As promised by the law of large numbers, as the data size $K^{sc}$ grows, $\bar{\mathbf{p}}^{ev}$ and $\Sigma$ converge to the true mean and true covariance matrix of $\mathbf{p}^{ev}$ \cite{zhang2018ambiguous}. Following the classic process of constructing a moment-based ambiguity set \cite{delage2010distributionally}, we propose the ambiguity set below: 
\begin{IEEEeqnarray}{l}
\mathcal{M}_{1}  = 
\Bigg\{ \mathbb{P}\in\mathcal{P}\big(\mathbb{R}^{|N_{evcs}|}\big) :  \notag\\
 \quad
\big(\mathbb{E}_{\mathbb{P}}[\mathbf{p}_k^{ev}] - \bar{\mathbf{p}}^{ev}\big)^{T}
\boldsymbol\Sigma^{-1}
\big(\mathbb{E}_{\mathbb{P}}[\mathbf{p}_k^{ev}] - \bar{\mathbf{p}}^{ev}\big)
\le \gamma_{1}, \notag \\
 \quad
\mathbb{E}_{\mathbb{P}}\Big[\big(\mathbf{p}_k^{ev}-\bar{\mathbf{p}}^{ev}\big)
\big(\mathbf{p}_k^{ev}-\bar{\mathbf{p}}^{ev}\big)^{T}\Big]
\preceq \gamma_{2}\,\boldsymbol\Sigma
\vphantom{\big(\mathbb{E}_{\mathbb{P}}[\mathbf{p}_k^{ev}]} \Bigg\}
\IEEEyesnumber \label{eq:amb3}
\end{IEEEeqnarray}
where $\mathcal{P}(\mathbb{R}^{|N_{evcs}|})$ represents the set of all probability distributions on $\mathbb{R}^{|N_{evcs}|}$. The first condition indicates that the true mean of the charging demand is within an ellipsoid of size $\gamma_1$ of the empirical mean, while the second bounds the covariance by $\gamma_2 \boldsymbol\Sigma$. Parameters $\gamma_1 \ge 0$ and $\gamma_2 \ge 1$ reflect the confidence level in empirical estimates and can be tuned via cross-validation.

In computing the empirical mean and covariance, average charge per EV is considered after rerouting. However, EV owners are sensitive to charging prices, i.e., charging less when prices are high and more when prices are low and this sensitivity is heterogenous. To capture this effect, a linear relationship between price difference and change in charging quantity is assumed:
\begin{IEEEeqnarray}{l}
\bar{n}_i = \frac{1}{K^{sc}} \sum_{k=1}^{K^{sc}} n_{i,k}, \quad \forall i \in N_{evcs} \IEEEyesnumber \IEEEyessubnumber* \label{eq:amb4a}\\
\bar{p}_i^{ev}(\lambda_i) = \sum_{n=1}^{\bar{n}_i} [Ch_{avg} - \beta (\Delta \lambda_i)], \forall i \in N_{evcs} \IEEEeqnarraynumspace \label{eq:amb4b}
\end{IEEEeqnarray}
where $\Delta \lambda_i$ is the price difference and $\beta$ is the sensitivity of an EV owner towards it. To this end, we establish an ambiguity set $\mathcal{M_{\mathrm{2}}}$ in equation \eqref{eq:amb5}. The ambiguity set incorporates second-order moment information to capture the stochastic dependence between EVCD and the charging price $\lambda$.   
\begin{IEEEeqnarray}{l}
\mathcal{M}_{2}  = 
\Bigg\{ \mathbb{P}\in\mathcal{P}\big(\mathbb{R}^{|N_{evcs}|}\big) :  \notag\\
 \quad
\big(\mathbb{E}_{\mathbb{P}}[\mathbf{p}_k^{ev}] - \bar{\mathbf{p}}^{ev}(\boldsymbol\lambda)\big)^{T}
\boldsymbol\Sigma^{-1}
\big(\mathbb{E}_{\mathbb{P}}[\mathbf{p}_k^{ev}] - \bar{\mathbf{p}}^{ev}(\boldsymbol\lambda)\big)
\le \gamma_{1}, \notag \\
 \quad
\mathbb{E}_{\mathbb{P}}\Big[\big(\mathbf{p}_k^{ev}-\bar{\mathbf{p}}^{ev}(\boldsymbol\lambda)\big)
\big(\mathbf{p}_k^{ev}-\bar{\mathbf{p}}^{ev}(\boldsymbol\lambda)\big)^{T}\Big]
\preceq \gamma_{2}\,\boldsymbol\Sigma
\vphantom{\big(\mathbb{E}_{\mathbb{P}}[\mathbf{p}_k^{ev}]}\Bigg\}
\IEEEyesnumber \label{eq:amb5}
\end{IEEEeqnarray}

\section{Proposed DOPF Formulation under Endogenous Uncertainty of EVCDs}
Currently, DSOs predict EVCD in their ahead-of-real-time decision-making. However, precisely predicting the EVCDs is extremely difficult since EV owners do not always behave rationally and prices are uncertain, this creates endogenous uncertainty that most existing models fail to capture. This study aims to minimize total system cost while ensuring feasibility via solving a DOPF problem under endogenous uncertainty.

\subsection{DOPF of Distribution System under Uncertainty}
With the power flow captured by the Distflow model \cite{li2016convex}, the DOPF problem under EVCD uncertainty can be written as follows: 
\begin{IEEEeqnarray}{rCl}
    \min_{p_i^G,q_i^G,v_i,p_{ik},q_{ik},\alpha_i} \sum_{i \in G} \mathbb{E}[C_i^G(\tilde{p}_i^G)] \IEEEyesnumber\IEEEyessubnumber* \label{eq:pdn4a} \\ 
    v_i - v_k - 2(r_{ik} p_{ik} + x_{ik} q_{ik})  = 0 \label{eq:pdn4b} \\ [3pt]
    p_i^{\mathrm{G}} - p_i^{load} - p_i^{ev} = \sum_k p_{ik} - \sum_j p_{ji} \label{eq:pdn4c} \IEEEyessubnumber* \\
    q_i^{\mathrm{G}} - q_i^{load} - zp_i^{ev} = \sum_k q_{ik} - \sum_j q_{ji}  \label{eq:pdn4d} \\
    \mathbb{P} (\mathbf{\mathcal{Y}}_i^{min} \leq \tilde{\mathbf{\mathcal{Y}}}_i \leq \mathbf{\mathcal{Y}}_i^{max}) \geq 1-\epsilon \label{eq:pdn4e} \\ [3pt]
    \mathbb{P}(\tilde{p}_{ik}^2 + \tilde{q}_{ik}^2 \leq S_{ik}^2) \geq 1-\epsilon \label{eq:pdn4f}
\end{IEEEeqnarray}
where $p_i^G, q_i^G$ represents the real and reactive power generation of controllable distributed energy resources at \textit{i}th node (the root node is also considered a generator), $p_i^{load}, q_i^{load}$ indicates real and reactive non EV loads at node \textit{i}, and  $p_{ik}, q_{ik}$ denotes the real and reactive  line flows at line \textit{ik}. Square of voltage magnitude is denoted by $v_i$ and $S_{ik}$ indicates the maximum allowable line flow. The violation parameter is $\epsilon$ and another parameter $z = tan\theta$ represents the reactive-to-real power ratio of EVCD which is assumed to be a fixed value. A vector $\mathbf{\mathcal{Y}}_i= [p_i^G, q_i^G, v_i]$ is declared to show the chance constraints. $\tilde{\mathbf{\mathcal{Y}}}_i, \tilde{p}_{ik}, \tilde{q}_{ik}$ denotes the respective uncertain decision variables.

The uncertain EVCD at node $i$ of the PDS can be written as
\begin{IEEEeqnarray}{rCl}
    \tilde{p}_i^{ev} = p_i^{ev} + \omega, \quad \mathbb{E}[\omega] = 0, \quad \text{Cov}[\omega] = \Sigma, \IEEEyessubnumber* \label{eq:pdn7}
\end{IEEEeqnarray}
where $p_i^{ev}$ denotes the predicted value of EVCD and $\omega$ is the uncertain variable with zero mean and covariance matrix $\Sigma$. The uncertainty influences the decision variables in (6), which are therefore expressed as functions of $\omega$. To demonstrate the generators' compensation method for the net load deviation, we employ the widely used affine real-time control:
\begin{IEEEeqnarray}{rCl}
\tilde{p}_i^G &=& p_i^G + \alpha_i \sum_j \omega_j, \qquad
\sum_i \alpha_i = 1, \IEEEyessubnumber*  \label{eq:pdn8}
\end{IEEEeqnarray}
where $\alpha_i$ allocates load deviations among controllable generators.

For radial networks, the matrix $\mathbf{A} \in \{0,1\}^{l \times b}$, where $l:= |L|$ is the number of lines and $b := |N_B^+|$ is the number of nodes except the root node \cite{mieth2018data}, is employed to map the change of EVCD at every node to the change of line power flow. Each line carries the net load of its downstream nodes. Using $A$ and power transfer distribution factor concepts, the uncertain line flows and voltages are 
\begin{IEEEeqnarray}{rCl}
    \tilde{p}_{ik} & = & p_{ik} + \sum_{j=1}^b \mathbf{A}_{ik,j} (\omega_j - \alpha_j \sum \omega_j) \IEEEyessubnumber*  \label{eq:pdn9} \\
    \tilde{q}_{ik} & = & q_{ik} + \sum_{j=1}^b \mathbf{A}_{ik,j} (\omega_j - \alpha_j \sum \omega_j)  \label{eq:pdn10} \\
    \tilde{v}_i & = & v_i - 2 \sum_{j=1}^l \mathbf{A}_{ij} \Bigg[\mathbf{R}_j \sum_{k=1}^b \mathbf{A}_{jk} (\omega_k - \alpha_k \sum \omega_k) \notag \\ 
    & & \qquad + \mathbf{X}_j \sum_{k=1}^b \mathbf{A}_{jk} (\omega_k - \alpha_k \sum \omega_k) \Bigg]  \label{eq:pdn11} 
\end{IEEEeqnarray}
where, $\textbf{R}, \textbf{X}$ are vectors of line resistance and reactances respectively.

\subsection{Deterministic Reformulation }
The DOPF formulation in (6) cannot be directly solved using mature solvers and therefore requires a deterministic reformulation. The chance constraints in \eqref{eq:pdn4e} and \eqref{eq:pdn4f} are reformulated into a deterministic equivalent to make them computationally tractable that can be easily solved with any off-the-shelf solver. Assuming a quadratic cost function for the controllable generators, under zero-mean forecast error and covariance $\Sigma$ of EVCS load, the expected generation cost in \eqref{eq:pdn4a} becomes:
\begin{IEEEeqnarray}{rCl}
    \mathbb{E}[C_i^G(\tilde{p}_i^G)] &=& \sum_{i \in G} [c_{i2}(p_i^{G})^2 + c_{i2}^{'} \alpha_i^2 + c_{i1} p_i^{G} + c_{i0}] \IEEEyesnumber \IEEEyessubnumber* \label{eq:rfm6a}
    \end{IEEEeqnarray}
where
\begin{IEEEeqnarray}{rCl}
    c'_{i2} &=& [\sum_{j=1}^{N^+} \sum_{k=1}^{N^+} \Sigma_{jk}] \times \alpha_i, i \in N_G \IEEEyessubnumber* \label{eq:rfm6b}
\end{IEEEeqnarray}

Here, $N_G$ is the number of generators. The chance constraints in \eqref{eq:pdn4e} can be reformulated in a second order conic form as follows:
\begin{IEEEeqnarray}{rCl}
    \mathbf{\mathcal{Y}}_i + \eta(\epsilon)(var(\tilde{\mathbf{\mathcal{Y}}}_i)^{1/2} \leq \mathbf{\mathcal{Y}}_i^{max} \IEEEyesnumber \IEEEyessubnumber* \label{eq:rfm6c} \\
    \mathbf{\mathcal{Y}}_i - \eta(\epsilon) (var(\tilde{\mathbf{\mathcal{Y}}}_i))^{1/2} \geq \mathbf{\mathcal{Y}}_i^{min} \label{eq:rfm6d}
\end{IEEEeqnarray}
where $var(\tilde{\mathbf{\mathcal{Y}}}_i)$ is calculated similarly as \cite{mieth2018data} and following \cite{zhang2018ambiguous} $\eta(\epsilon)$ is calculated as
\begin{IEEEeqnarray}{rCl}
    \eta(\epsilon) = \begin{cases}
    \sqrt{\gamma_1} + \sqrt{(\frac{1-\epsilon}{\epsilon})(\gamma_2 - \gamma_1)}, & \quad \text{if} \ \frac{\gamma_1}{\gamma_2} \leq \epsilon \\
    \sqrt{\frac{\gamma_2}{\epsilon}}, & \quad \text{if} \ \frac{\gamma_1}{\gamma_2} > \epsilon
    \end{cases} \IEEEyessubnumber* \label{eq:rfm6e} 
\end{IEEEeqnarray}

The quadratic line limit constraint is first approximated by a 12-edge polygon. After that, conic reformulation is used to convert it to a tractable form.
\begin{IEEEeqnarray}{l}
    b_{1,c} \tilde{p}_{ik} + b_{2,c} \tilde{q}_{ik} - b_{3,c} S_{ik} \le 0, c=1,\dots,12, \forall ik \in L \IEEEeqnarraynumspace \IEEEyesnumber \IEEEyessubnumber* \label{eq:rfm6f}
\end{IEEEeqnarray}
Applying conic reformulation:
\begin{IEEEeqnarray}{rCl}
b_{1,c} [p_{ik} &+& \eta(\epsilon) (var(\tilde{p}_{ik}))^{1/2}] + b_{2,c} [q_{ik} + \eta(\epsilon) (var(\tilde{q}_{ik}))^{1/2}] \notag \\ 
&& \qquad - b_{3,c} S_{ik} \leq 0, \quad \forall ik \in L \IEEEyessubnumber* \label{eq:rfm6g}
\end{IEEEeqnarray}
where, $\eta(\epsilon)$ is the same as \eqref{eq:rfm6e} and $var(\tilde{p}_{ik})$ and $var(\tilde{q}_{ik})$ are calculated similarly as \cite{mieth2018data}.

\section{Case Study and Test Results}
\subsection{Case Study}
The proposed PROMETHEE-based framework is tested on IEEE 33-bus and 123-bus distribution systems. We assume the DSO has access to EVCS data (number of EVs at a particular time of the day, congestion at EVCS). The output of the PM model is subsequently used to construct the ambiguity set $\mathcal{M_{\mathrm{2}}}$. The parameters $\gamma_{1}$ and $\gamma_{2}$ are determined through cross-validation, which yields $\gamma_{1} \in [0.01,0.2]$ and $\gamma_{2} \in [1,2]$. For the case studies, $\gamma_{1}$ and $\gamma_{2}$ are set to 0.1 and 1, respectively, to ensure robustness of the model. The violation parameter $\epsilon=0.1$, i.e., confidence level is $90\%$. We then use the first and second order moment from the ambiguity set to reformulate the chance constraints of the DOPF model and solve the problem.

\subsubsection*{Case A}
We use modified IEEE 33-node distribution system for the test with four EVCS in the system. Each EVCS is connected to a distinct feeder, and they are placed at different nodes of different feeder. The model's performance with different EVCD percentage with respect to non-EV loads is shown in Table~\ref{tab:impact}.   

\subsubsection*{Case B}
To evaluate the model under a more complex network topology, second test is conducted on a modified IEEE 123-node distribution system. Similar to Case A, four EVCSs are considered, each served by a separate feeder. However, the EVCSs are located at different nodes within their respective feeders. The impact of different EVCD penetration levels on system performance is presented in Table~\ref{tab:impact}.   

\begin{table}[!t]
  \caption{Impact of EVCD Penetration}
  \label{tab:impact}
  \centering
  \footnotesize
  \resizebox{\columnwidth}{!}{%
  \begin{tabular}{lcccc}
    \toprule
    \textbf{System} & \textbf{EVCD (\%)} & \textbf{EVCD (MW)} & \textbf{Total Load (MW)} & \textbf{Cost (\$)} \\ 
    \midrule
    \multirow{4}{*}{33 node} 
      & 0.00  & 0.00 & 29.72 & 11894.62 \\ 
      & 4.44  & 1.32 & 31.04 & 12464.47 \\ 
      & 10.13 & 3.01 & 32.73 & 13236.81 \\ 
      & 15.56 & 4.63 & 34.35 & 14028.57 \\ 
    \midrule
    \multirow{4}{*}{123 node} 
      & 0.00  & 0.00 & 27.92 & 11173.84 \\ 
      & 7.05  & 1.97 & 29.89 & 11962.34 \\ 
      & 19.65 & 5.48 & 33.40 & 13405.70 \\ 
      & 25.36 & 7.08 & 35.00 & 14110.37 \\ 
    \bottomrule
  \end{tabular}
  }
\end{table}

\subsection{Post Optimization Analysis}
\begin{table}[!t]
  \caption{Performance of Post Optimization Analysis}
  \label{tab:performance}
  \centering
  \footnotesize
  \resizebox{\columnwidth}{!}{%
  \begin{tabular}{llccc}
    \toprule
    \textbf{System} & \textbf{Benchmark} & \textbf{Deterministic} & \textbf{DRCC} & \textbf{DRCC+PM} \\
    \midrule
    \multirow{4}{*}{33 node} 
     & Average cost & 271775.85 & 241819.21 & 191963.14 \\
     & Cost Increase (\%) & 19.96 & 6.73 & - \\
     & Imbalance metric $\bar{D}$ & 5.33 & 4.35 & 4.06 \\
     & Constraint violation (\%) & 5.6 & 3.9 & 3.5 \\
    \midrule
    \multirow{4}{*}{123 node} 
     & Average cost & 443604.10 & 439196.54 & 364197.06 \\
     & Cost Increase (\%) & 21.80 & 20.59 & -  \\
     & Imbalance metric $\bar{D}$ & 7.68 & 7.34 & 6.42 \\
     & Constraint violation (\%) & 14.7 & 14.3 & 10.7 \\
    \bottomrule
  \end{tabular}
  }
\end{table}
In uncertainty-aware optimization frameworks \cite{li2021uncertainty}, post-optimization analysis is essential to evaluate the robustness and feasibility of dispatch decisions under realistic realizations of uncertain parameters. To evaluate the effectiveness of the proposed PROMETHEE-based DRCC model, a post-optimization analysis is conducted. This analysis compares three approaches: the deterministic model, the standard DRCC model without PROMETHEE, and the PROMETHEE-enhanced DRCC model (DRCC+PM). For each model, 1,000 realizations of EV charging demand are generated at each EVCS as the testing scenarios. The dispatch decisions obtained from each model are then re-evaluated under these realizations using a slack-based redispatch formulation that penalizes violations in power balance, voltage limits, and line capacity constraints.

In this redispatch framework, all decision variables are fixed to their original values from the respective models, except for slack variables that absorb any infeasibility. The DRCC and DRCC+PM models use the $\alpha$ obtained directly from the optimization. For comparison with the deterministic model, we assume that deterministic model applies a uniform policy, assuming equal generator participation, i.e., $\alpha_{i} = 1/N_{G}$ \cite{lubin2019chance}.

Now, we want to compare feasibility of the obtained solutions of Deterministic, DRCC and DRCC+PM models. If the slack variables for real power redispatch, $\bar{s}_i^p$ and $\underline{s}_i^p$ are non-zero that means the original solution cannot fully counteract the uncertainty realization. Therefore, to check this infeasibility, imbalance metric $ \bar{D}= \sum_{i \in G} \bar{s}_i^p $ and $\underline{D}= \sum_{i \in G} \underline{s}_i^p $ are computed for each realization \cite{lubin2019chance}. 
\begin{figure}[!t]
    \centering
    \includegraphics[width=1\columnwidth, height= 0.20\textheight]{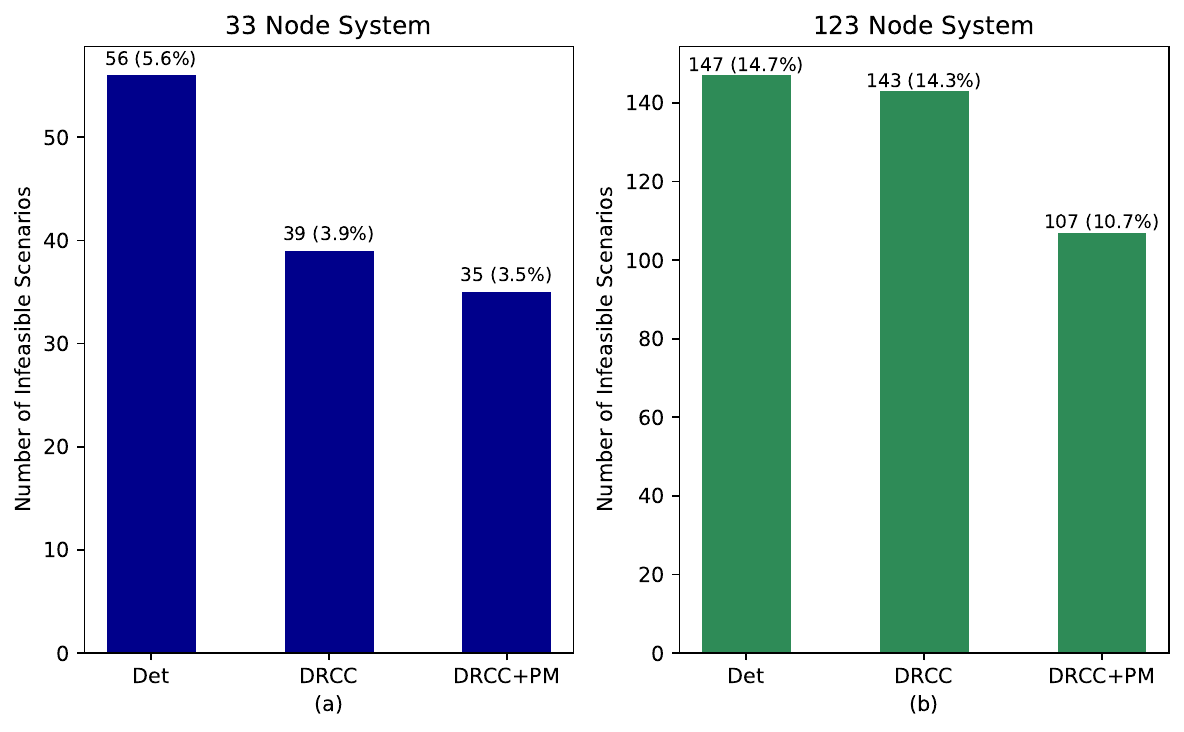}
    \vspace{-10pt}
    \caption{Constraint violation percentages for different models of (a) 33 node system (b) 123 node system}
    \label{fig:2}
\end{figure}

Table \ref{tab:performance} along with the constraint violation analysis, highlight the superior performance of the proposed DRCC+PM model. In the 33-node system, DRCC+PM achieves the lowest post-optimization cost—over 6\% lower than the optimized DRCC model and nearly 20\% lower than the deterministic baseline. The 123-node system shows similar trends, with DRCC+PM  yielding the least post-optimization cost. The DRCC+PM model also yields the lowest imbalance metric out of the three models for both the systems. Constraint violation analysis reveals that DRCC+PM has the lowest violation of constraints in both the 33-node and 123-node systems.

\section{Conclusion And Future Work}
This paper proposed a novel framework for modeling EVCD that explicitly captures the heterogeneous preferences of EV owners and the endogenous uncertainty arising from their charging behavior. By employing PROMETHEE within a DRCC (DRCC+PM) formulation, the proposed approach systematically integrates both rational factors and psychological factors into demand modeling. Comprehensive case studies on the IEEE 33-node and 123-node distribution systems demonstrate the effectiveness of the proposed DRCC+PM framework. Comparative analyses against deterministic and standard DRCC formulations reveal that the DRCC+PM approach consistently achieves superior performance. These findings confirm that explicitly incorporating behavioral uncertainty into EVCD modeling enables DSOs to more accurately capture demand variability, allocate resources more effectively, and mitigate risks of congestion and instability in distribution networks. Future work will focus on extending the proposed framework to incorporate transportation system dynamics and on validating the approach using large-scale real-world EV datasets to assess its scalability and practical deployment potential.

\bibliographystyle{IEEEtran}
\bibliography{references}

\end{document}